\documentclass[prl,twocolumn,preprintnumbers,amsmath,amssymb,floatfix,aps,10pt]{revtex4-1}
\usepackage{amsmath,cleveref,graphicx,bm,nicefrac,color,subfigure}
\newcommand{\e}{\text{e}}
\newcommand{\ASlash}{/ \!\!\!\! A}
\newcommand{\kSlash}{/ \!\!\! k}

\newcommand{\epsSlash}{/ \!\!\! \epsilon}
\newcommand{\Felix}[1]{\textcolor{black}{#1}}

\begin{document}
\title{Nonlinear Compton scattering of an ultra-intense laser pulse in a plasma}
\author{Felix Mackenroth}
\email{mafelix@pks.mpg.de}
\affiliation{Max-Planck-Institut f\"ur Physik komplexer Systeme, N\"othnitzer Str. 38, 01187 Dresden, Germany}
\affiliation{Max-Planck-Institut f\"ur Kernphysik, Saupfercheckweg 1, 69117 Heidelberg, Germany}
\author{Naveen Kumar}
\email{kumar@mpi-hd.mpg.de}
\author{Norman Neitz}
\author{Christoph H. Keitel}
\affiliation{Max-Planck-Institut f\"ur Kernphysik, Saupfercheckweg 1, 69117 Heidelberg, Germany}
\date{\today}

\begin{abstract}
\Felix{Laser pulses traveling through a plasma can feature group velocities significantly differing from the speed of light in vacuum.} This modifies the well-known Volkov states of an electron inside a strong laser-field from the vacuum case and consequently all quantum electrodynamical effects triggered by the electron. Here we present an in-depth study of the basic process of photon emission by an electron scattered from an intense short laser pulse inside a plasma, labeled nonlinear Compton scattering, based on modified Volkov solutions derived from first principles. \Felix{Consequences of the nonlinear, plasma-dressed laser dispersion on the Compton spectra of emitted photons and implications for high-intensity laser-plasma experiments are pointed out.}
\end{abstract}

\maketitle

\allowdisplaybreaks

\Felix{Exposing matter to lasers of highest electromagnetic field strengths~\cite{DiPiazza_etal_2012} holds promises ranging from technological applications such as compact particle accelerators~\cite{Esarey_etal_2009,Daido_etal_2012,Macchi_etal_2013} or radiation sources~\cite{Chen_Maksimchuk_Umstadter_1998,TaPhuoc_etal_2003,Schwoerer_etal_2006,TaPhuoc_etal_2012,Huang_etal_2012,Corde_etal_2013,Sarri_etal_2014,Powers_etal_2014} to groundbreaking studies of nonlinear quantum electrodynamics (QED) effects. In vacuum, the scattering of an electron (mass $m$ and charge $e<0$) with initial four-momentum $p^\mu = \varepsilon(1,\beta \bm{n})/c$, where $\beta$ is the electron's velocity in units of the speed of light in vacuum $c$, from a plane wave with electric field amplitude $E$ and wave vector $k_L^\mu = \omega_L(1,\bm{n}_L)/c$ becomes nonlinear in the regime $\xi = |e|E/c m\omega_L\gtrsim1$, and quantum effects dominate for $\chi = ((p_\mu k_0^\mu)/mc\omega_L) E/E_\text{cr}\gtrsim1$, where $E_\text{cr} = m^2 c^3/\hbar\left|e\right|$ is the critical field of QED \cite{Ritus_1985}. Such nonlinear QED effects are conventionally accounted for by including them in numerical simulations of laser-matter interactions. While many theoretical efforts continue to further improve these QED-laser-plasma simulation schemes~\cite{Ridgers_etal_2014,Gonoskov_etal_2015,Grismayer_etal_2016}, they are all based on approximating the QED rates by incoherent single-particle rates in a constant plane wave, since for high particle energies and laser intensities any electromagnetic field can be approximated by a plane wave, constant on the short time scales of QED processes \cite{Ritus_1985,Baier_b_1998}. The corresponding QED calculations, however, assume the plane wave to propagate through vacuum, i.e., its group velocity to be $c$. In a realistic experiment on ultra-intense laser-matter interaction, on the other hand, there will be a background of massive particles present, quickly ionized to a plasma of density $n_e$, introducing the plasma frequency $\omega_p =\sqrt{4\pi n_e e^2 /m \gamma_L}$ with $\gamma_L=\sqrt{1+\xi^2/2}$ as new time scale. In underdense plasmas $\omega_p\leq\omega_L$ ultra-short laser pulses propagate over macroscopic distances \cite{Esarey_etal_2009} almost entirely within the plasma and trigger rich dynamics in it only on time scales longer than the pulse duration~\cite{Gorbunov_etal_2005,Lu_etal_2007}. Depending on the plasma's parameters, however, its presence may lead to profound alterations of the QED phenomenology, e.g., the photons' dispersion relation was shown to exhibit features reminiscent of the Higgs mechanism \cite{Anderson_1963}, and the vacuum approximation becomes invalid.} 
\Felix{The most basic nonlinear QED effect conventionally considered in ultra-intense laser-matter interactions is the emission of quantized radiation, or Compton scattering. Linear Compton scattering, i.e., the absorption and emission of only one photon by an electron has been studied in the presence of a plasma background already some time ago~\cite{Blumenthal_Gould_1970,Goldstein_Lenchek_1971}. However, novel laser pulses feature unprecedentedly high photon fluxes ($\xi\gg1$), leading to an altered effect, labeled nonlinear Compton scattering, commonly approximated as an incoherent sequence of emissions of single high-energy photons upon coherent absorption of photon from a laser propagating through vacuum~\cite{DiPiazza_Hatsagortsyan_Keitel_2010}. This process can no longer be described by conventional linear QED, indicating the need for a nonlinear theory of laser-electron interactions. In the presence of a background plasma such a theory is thus far missing. On the other hand, the importance of understanding the emission originating from the interaction of intense laser pulses with the electron population of a dilute plasma is also signified by earlier, classical analyses of nonlinear laser-driven particle dynamics~\cite{Akhiezer_Polovin_1956}, later refined to also include the emitted power spectrum~\cite{Waltz_Manley_1978}. Preliminary studies of this influence were performed at low laser intensities in a classical framework~\cite{Castillo-Herrera_Johnston_1993}, similarly to studies of laser-driven electrons in vacuum~\cite{Sarachik_Schappert_1970,Salamin_Faisal_1996}, however with no QED effects taken into account.}

In this Letter, \Felix{we provide a first-order or leading-order analysis of the impact of a nonlinear photon dispersion on nonlinear Compton scattering in a QED framework. This study's main goal is to investigate the validity of the hitherto used QED phenomenology model based on scatterings in vacuum. Due to the complexity of the underlying nonperturbative QED framework, we perform a leading-order analysis of the plasma's influence. By identifying deviations of the corresponding perturbative plasma effects from the vacuum theory, we put bounds on the latter's applicability. We thus do not take into account detailed plasma \textcolor{black}{dynamics such as instability excitations}. Much rather, we make the simplifying assumption of the background plasma to be cold and collisionless, which was demonstrated to be reasonable for relativistic laser-matter interactions~\cite{Kruer_2003} provided ion motion is negligible and the electron temperature $T_e$ is small compared to the energy gain from the laser field $m\xi \gg T_e$} \cite{Feit_Garrison_Rubenchik_1996}. The plasma ions (mass $m_i\gg m$), however, can be heated only on \textcolor{black}{time scales corresponding to several ion plasma periods} \textcolor{black}{$t_i =2\pi/ \omega_{p,i}^{-1} =\omega_p^{-1}(m\to m_i)\gg \omega_p^{-1}$} which we ensured to be much longer than the laser pulse durations considered in our work, indicating that the ions also remain cold. Consequently, the plasma's effect on the particle dynamics is negligible and its impact on the electromagnetic fields can be treated by a mean-field approximation, yielding a nonlinear dispersion relation. In accordance with most ultra-high field facilities operating in the optical regime, we assume the laser pulse to have an optical carrier frequency, whence it will experience the background plasma as a refractive index \Felix{$\rho\equiv\rho(\omega_L)=(1-(\omega_p/\omega_L)^2)^{1/2}\neq1$ resulting in a changed wave vector $k_L^\mu = \omega_L(1,\rho\bm{n}_L)$ with $k_L^2\neq0$} \Felix{\cite{Piel_2014}}. Since the electron emits mainly high-energy harmonics of its base frequency, on the other hand, we may assume the emitted photons to be unaffected by the background plasma. The basis for nonperturbative QED are solutions of the Dirac equation in the field under consideration, which, for a laser propagating through a medium with nonlinear dispersion relation, have been a long-standing issue and several solutions were communicated~\cite{Becker_1977,Cronstroem_Noga_1977,Mendonca_Serbeto_2011,Raicher_Eliezer_2013,Varro_2014,Heinzl_Ilderton_King_2016}, discriminating between lightlike and spacelike photon fields. Many of these solutions, however, were of exploratory nature~\cite{Bergou_Varro_1980,Varro_2013} until recently a more quantitative study was put forward~\cite{Heinzl_Ilderton_King_2016}. It was shown that in scattering problems involving energy scales far above any binding barrier, such as studied here, a perturbative approach for the wave function is satisfactory. However, transitions between these electron states have not yet been investigated. In QED such transitions are mediated by the emission of photons, and have been analyzed for special non-plane wave geometries recently~\cite{Raicher_Eliezer_Zigler_2014,Raicher_Eliezer_Zigler_2015a,Raicher_Eliezer_Zigler_2015b,Raicher_Eliezer_Zigler_2016}, carrying some resemblance to the full problem of Compton scattering in a plasma background.

We thus start our analysis from the Dirac equation in the presence of a strong, plane wave laser pulse of amplitude $A_L^\mu (\eta):= A_L^\mu\, \psi(\eta) = \epsilon_L^\mu \psi(\eta) ({m\xi}{/|e|})$ with an optical carrier frequency $\omega_L = 1.55$ eV (units $\hbar = c = 1$ are used) where $\eta=k_L^\mu x_\mu$ is the invariant laser phase \Felix{and $\psi(\eta)$ the potential's shape function}. Based on multiple scale perturbation theory~\cite{Bender_MathMethods} we found the following solutions to this equation (s.~\cite{Heinzl_Ilderton_King_2016} for an analogous derivation)
\begin{align}
\Psi_p(x) &= \bigg[\Phi_{V,p} + \frac{k_L^2}{2(k_Lp)} \delta \Phi_p \bigg]\e^{-i px + {i\Sigma(\eta)}} \frac{u_p}{\sqrt{2 \varepsilon V}} \label{Eq:WaveFunction}\\
{\Sigma(\eta)} &= \int_{-\infty}^\eta d\phi \sigma_p(\phi) + \frac{k_L^2}{2(k_Lp)} \sigma_p^2(\phi) \nonumber \\
\delta \Phi_p &= \sigma_p(\eta)\left\{1+e \frac{\kSlash_L \ASlash_L (\eta)}{(k_Lp)}\right\}\nonumber \\ &- \frac{e^2A_L^2(\eta)}{4(k_Lp)}\Phi_{V,p} - i e \frac{\kSlash_L \ASlash_L'(\eta)}{2(k_Lp)}, \nonumber
\end{align}
where {$\Phi_{V,p} = 1 + e \kSlash_L \ASlash_L (\eta)/2(k_Lp)$}, $u_p$ is the bi-spinor for a free electron of momentum $p^\mu$, the prime denotes differentiation with respect to $\eta$, $\sigma_p(\eta) =  e^2A^2_L(\eta)/2 (k_Lp) - e (pA_L(\eta))/(k_Lp)$ and the Feynman slash notation $\not\!\!a = \gamma^\mu a_\mu$ with the Dirac matrices $\gamma^\mu$ is used. It can be shown that this above solution is formally equivalent to the solution resulting from a perturbative expansion of the full Dirac equation in orders of $k_L^2$. We then use these wave functions (\ref{Eq:WaveFunction}) as basis set for a strong field expansion of QED in a laser field propagating through a plasma to compute the probability of the emission of a photon with wave vector $k_1^\mu=\omega_1(1,\bm{n}_1)$ towards $\bm{n}_1=(\cos(\phi_1)\sin(\theta_1),\sin(\phi_1)\sin(\theta_1),\cos(\theta_1))$ from an electron changing its initial momentum $p_i^\mu$ to a final momentum $p_f^\mu$. The scattering matrix element of this process is given by \cite{Ritus_1985,Mackenroth_2014}
\begin{align}\label{Eq:ScatteringMatrixElement}
 S_{fi} = -ie \sqrt{\frac{4\pi}{2\omega_1V}} \int d^4x \overline{\Psi}_{p_f}(x) \epsSlash_1^* \e^{ik_1x} \Psi_{p_i}(x), 
\end{align}
where $\overline{\Psi}_p(x) = \Psi^*_p(x)\gamma_0$ is the wave function's Dirac conjugate and ${\epsilon_1^\mu}^*$ the emitted photon's polarization vector's complex conjugate. We note that three of the four space-time integrals in this expression yield energy-momentum conservation like in the vacuum analysis~\cite{Boca_Florescu_2009,Mackenroth_DiPiazza_2011,Seipt_Kaempfer_2011,Krajewska_Kaminski_2012_a} while, unlike in the vacuum case, the nontrivial integration is in the plasma dressed laser phase $\eta = \omega_L(t-\rho\, \bm{n}_L\bm{x})$. This phase variable will no longer be a Lorentz invariant, as the refractive index depends on the spatial plasma density, whence we limit our discussion to the experimentally most relevant reference frame in which the plasma is on average at rest. \Felix{From \cref{Eq:ScatteringMatrixElement} one can now obtain the emitted energy according to $d\mathcal{E}=\omega_1 d\Gamma_1d\Gamma_f \sum_{\sigma,\lambda}\left|S_{fi}\right|^2$ where $\sum_{\sigma,\lambda}$ indicates summing (averaging) over final (initial) state spins and polarizations and $d\Gamma_1=d\bm{k}_1/(2\pi)^3$ is the emitted photon's phase space and we use energy-momentum conservation to collapse the integrals over the final state electron's phase space $d\Gamma_f$. We ensured several analytical limits of the resulting QED radiation probability: We found that in the linear limit $\xi\to0$ the expression reduces to the perturbative QED amplitude of a dispersive photon undergoing Compton scattering off an electron. We also found that in the vacuum limit $n_e\to0$ it reduces to the well-known expressions of nonlinear Compton scattering in vacuum \cite{Boca_Florescu_2009,Mackenroth_DiPiazza_2011,Seipt_Kaempfer_2011,Krajewska_Kaminski_2012_a}. Finally, we also confirmed that in the classical limit $\chi\to0$ the QED current $j^\mu = \bar{\Psi}_p(x) \gamma^\mu \Psi_p(x)$ resulting from the used wave functions reduces to its classical counterpart. In order to obtain this classical current, necessary to compute classical emission spectra, we solved the classical equations of motion inside a laser field propagating through a background plasma, mediated by a modification of the laser photons' dispersion relation. To obtain this solution we observe that for $k_L^2\neq 0$ the classical equations of motion become (s. also \cite{Heinzl_Ilderton_King_2016})
\begin{align}
  \frac{d  p^\mu (\eta)}{d \eta} =\frac{e}{(p(s)k_L)}(k_L^\mu(A_Lp(s))-A_L^\mu(p(s)k_L))\partial_\eta\psi_\mathcal{A}(\eta).
\end{align}
\begin{figure}
\includegraphics[width=0.45\textwidth,height=0.35\textwidth]{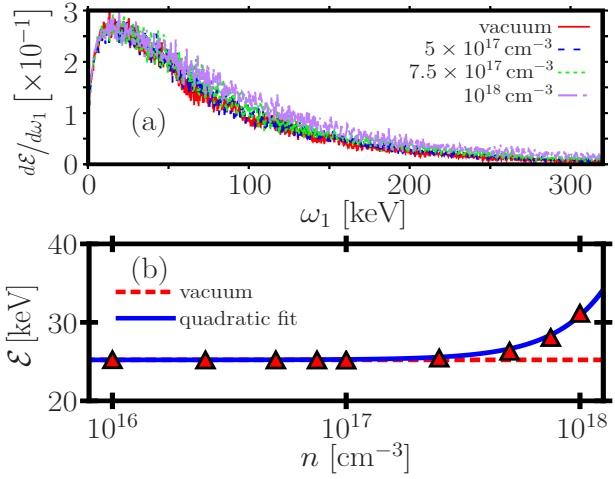}
\caption{(a) Integrated spectra for $I_L=10^{22}\, \text{W}/\text{cm}^2$ for different plasma densities. (b) Total emitted energy compared to the vacuum limit (dashed red) with a quadratic fit (solid blue).}
\label{fig1}
\end{figure}
This equation is a complete integral and solved by the electron momentum $p^\mu(\eta)=p_i^\mu-eA_L^\mu\psi_\mathcal{A}(\eta)+ k_L^\mu\left[(p(\eta)k_L)-(p_ik_L)\right]/k_L^2$. As a consequence, the emission probability obtained from the modulus square of \cref{Eq:ScatteringMatrixElement} agrees with the classical radiation power, obtained from inserting the above derived classical electron current into the Li\'enard-Wiechert potentials, up to terms of order $k_L^4$ in agreement with the here used order of the wave function's expansion.} There is, however, a discrepancy between the classical and quantum emission inside a plasma: The modulus square of the classical current, entering the emission probability, for an electron initially at rest in a plasma is corrected by $j^2_\text{plas} = j_\text{vac}^2 [1 - \xi^2 k_L^2/\omega_L^2 ]$. Consequently, we see that classically a plasma suppresses radiation. In the quantum case, e.g., for high emitted frequencies, the emission occurs close to the stationary points $\eta_0$ of the rapidly oscillating exponential phase, here distinguished by $\sigma_{p_i}(\eta_0) - \sigma_{p_f}(\eta_0) + k_L^2/2(\sigma_{p_i}^2(\eta_0)/(k_Lp_i) -\sigma_{p_f}^2(\eta_0)/(k_Lp_i)) = 0$. In the vacuum case for an electron initially at rest, as studied below, these stationary points are distinguished by the potential's shape function assuming a value $\psi(\eta_0^\text{vac}) = -e(p_fA)(k_Lp_i)/e^2A_L^2(k_Lk') + i \kappa$, where the imaginary part is small $\kappa \sim 1/\xi$ and the real part determines the angular emission's distribution~\cite{Mackenroth_etal_2010}. Inserting this solution back into the original equation we find the stationary points in the presence of a background plasma to be solutions of the equation $\psi(\eta_0^\text{plas}) = \psi(\eta_0^\text{vac}) + i \mathcal{C}$, with some complicated real factor $\mathcal{C}$, i.e., the plasma induces a purely imaginary correction to the stationary point and the angular emission range, distinguished by the latter's real part is expected to remain unchanged. On the other hand, close to the real stationary points the exponential phase is corrected in the background plasma by a factor $\delta \Sigma(\eta_0) \approx - k_L^2 (k_Lp_i)^3(p_fA)^4/4A_L^4(k_Lp_f)^3(k_Lk')^3 (1+(k_Lk_1)/(k_Lp_i)-(k_Lp_i)/(k_Lk_1))\Delta\eta_\text{coh} <0 $, where $\Delta\eta_\text{coh}$ indicates the process' coherence length. In the high-frequency regime, $(k_Lk_1)\sim (k_Lp_i)$ this correction is negative, i.e., the exponential oscillations are reduced and the radiation probability enhanced. Physically, this seems to imply that while classically the plasma suppresses the radiating charge current, if quantum effects are important the suppression concerns quantum phase oscillations, in fact enhancing the emission.
\begin{figure}
\includegraphics[width=\linewidth]{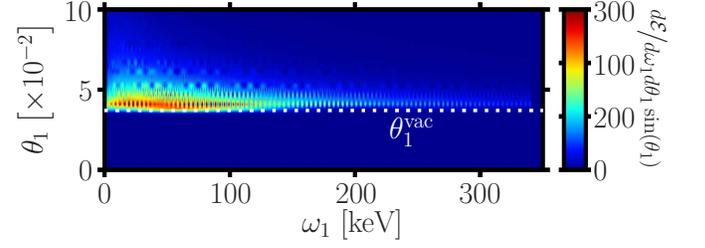}
\caption{\Felix{Angular spectra of the Compton scattered signal for a laser pulse with $I_L = 10^{22} \text{W}/\text{cm}^2$ ($\xi\approx70$) and a plasma density of $n_e=10^{18}$cm$^{-3}$ compared to the vacuum boundary angle $\theta_1^\text{vac}$ (white dotted line).}}
\label{fig2}
\end{figure}

\Felix{Next, we quantify the above discussion in the context of the ongoing and planned high-intensity laser-plasma interaction experiments~\cite{DiPiazza_etal_2012} by integrating \cref{Eq:ScatteringMatrixElement} numerically. We are going to study a two-cycle laser pulse with $\psi(\eta)=\sin^4(\eta/4)\sin(\eta)$ if $\eta\in[0,4\pi]$ and zero elsewhere scattering an electron initially at rest $p^\mu_i=(m,\bm{0})$. In accordance with the cold, collisionless plasma approximation any plasma electron can be approximated by this initial state, as its random thermal motion is negligible compared to its laser-driven dynamics. We begin by considering a laser of intensity $I_L = 10^{22} \text{W}/\text{cm}^2$ ($\xi\approx70$) and visualize the full radiation process by integrating the emission probability over all directions of the emitted photon to obtain the energy emitted per unit frequency $d\mathcal{E}/d\omega_1$ for different plasma densities (s.~Fig.~\ref{fig1} (a)). While the spectral peak at low $\omega_1$ is unaffected by the plasma, the collective effect of the optical photons accumulates into a higher yield of high-energy photons, even though high-energy photons do not see the plasma as a medium. Integrating the spectra over all frequency components, we obtain the total emitted energy $\mathcal{E}$ as a function of plasma density (s.~Fig.~\ref{fig1} (b)). The numerical data is well reproduced by a quadratic fit $\mathcal{E} = \mathcal{E}(n_e=0) + \delta \mathcal{E}\, n_e^2$, with $\delta \mathcal{E} \approx 6 \times 10^{-33} \text{ eV cm}^6$, demonstrating an increasing plasma density to lead to a nonlinear increase of emitted energy with respect to the vacuum result, which reduces to the Larmor formula for $\chi\to0$ \cite{Jackson_1999}. Next to this spectral analysis, the emission's angular distribution is of interest, which in vacuum was shown to be confined to $\theta_1\geq \theta_1^\text{vac} := 2\varepsilon/m\xi \psi^\text{max}$ \cite{Harvey_etal_2009,Mackenroth_etal_2010}, where $\psi^\text{max}$ is the shape function's maximal value, i.e., in the present case $\psi^\text{max}\approx0.78$. Integrating \cref{Eq:ScatteringMatrixElement} over $\phi_1$ we obtain the angular spectrum which even for the largest plasma density studied above exhibits the same angular confinement (s.~Fig.~\ref{fig2}).} Thus, the emission's angular distribution is not influenced by the background plasma, unlike its spectrum.
\begin{figure}
\includegraphics[width=\linewidth]{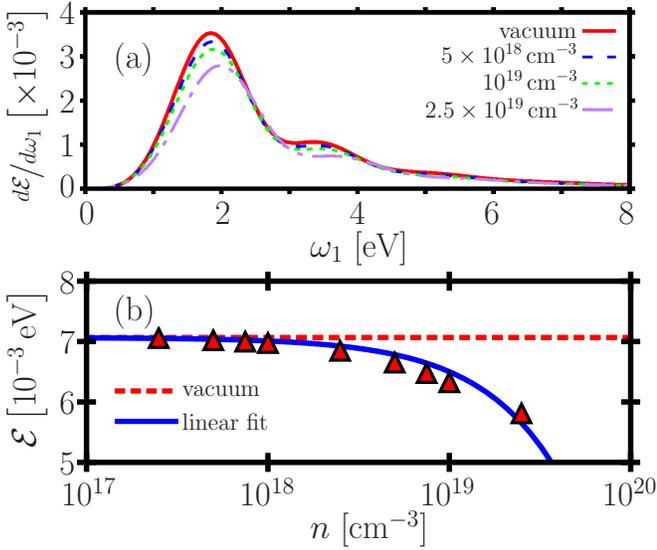}
\caption{(a) Integrated spectra for $I_L=10^{22}\, \text{W}/\text{cm}^2$ for forward emission ($\theta_1\leq\theta_1^\text{vac}/2$) for different plasma densities. (b) Total emitted energy compared to the vacuum limit (dashed red) with a linear fit (solid blue).}
\label{fig3}
\end{figure}

\textcolor{black}{Apart from observing the high-yields of the high-energy photons, one can also investigate the Compton scattering of low-energy photons. Though it may seem counterintuitive at first sight but it's important for two reasons: first, the high-energy photons do not experience the plasma's refractive index but low-energy photons do, satisfying a non-linear dispersion relation. Thus, the influence of the plasma, as a medium, on the Compton scattering can be different for low-energy photons. Second, the Compton scattered signal at low-energy photons can interfere with the stimulated Raman scattering (SRS) instability, which is the manifestation of the collective effects of the plasma medium and it's one of the most important parametric instabilities in a plasma~\cite{Barr_Mason_Parr_1999,Mori_etal_1994,Quesnel_etal_1997,Sakharov_Kirsanov_1997,Kumar_etal_2013}. Indeed, the non-resonant Raman scattering is often refereed to as the stimulated Compton scattering in a plasma ~\cite{Kruer_2003}. The SRS of a laser pulse can occur in any direction, but the forward Raman scattering (FRS) branch of the SRS propagates collinearly with the laser pulse, and affects the low-energy laser photons in the laser's propagation direction, i.e., inside the central dip of the emission's angular distribution (s. Fig.~\ref{fig2}). The SRS of a laser pulse is usually probed by interferometric analyses on the laser pulse itself~\cite{Everett_etal_1995} while the high-energy photons from Compton scattering are recorded on a detector. Thus, the detection of the Compton scattering of a laser pulse  in a plasma has twofold possibilities both in low-photon energy (eV) and high-photon energy (MeV) regimes. These detection possibilities, instead of being intrusive, are complementary to each other and their simultaneous observations can further affirm the theoretical predictions presented here.} 
\begin{figure}[t!]
\centering
\includegraphics[height=0.32\textwidth]{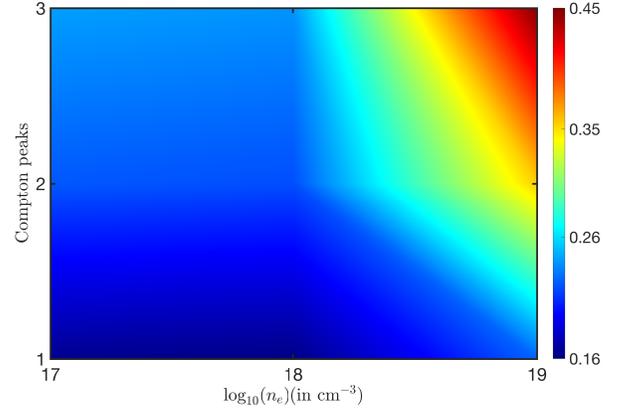}
\caption{Difference between Compton peaks corresponding to Fig.~\ref{fig1} and the harmonics of anti-Stokes mode $\omega_{+} = n \omega_L + \omega_p$, where $n=1,2,3$ (corresponding to different Compton peaks), for $\xi=100$ at different plasma densities. The colorbar represents the photon energies in eV.}
\label{fig4}
\end{figure}
%

\Felix{In order to \textcolor{black}{facilitate this comparison}, we integrate \cref{Eq:ScatteringMatrixElement} over all $\phi_1$ but only over $\theta_1\leq\theta_1^\text{vac}/2$. In this direction the integrated spectra feature only a few peaks (s. Fig.~\ref{fig3} (a)), similar to the vacuum case where for $\theta_1=0$ only the first harmonic $\omega_1\equiv\omega_L$ can be emitted. As argued above, this low-energy emission is reduced at higher plasma densities (s. Fig.~\ref{fig3} (b)) and the data is well reproduced by a linear fit $\mathcal{E} = \mathcal{E}(n_e=0) - \delta \mathcal{E}\, n_e$, with $\delta \mathcal{E} \approx 6\times10^{-23} \text{ eV cm}^3$, demonstrating that the reduction with respect to the vacuum Larmor result is due to linear plasma effects. Furthermore, the peak positions shift to larger $\omega_1$}. Since in a plasma $k_{L}^{2} = \omega^{2}_p$, one can interpret the effective laser photon energy in a plasma as $\omega_L^\text{plas} = \omega_L + \omega_p$, equivalent to the quantum of the anti-Stokes mode of the Raman scattering. Thus one can compare the emission frequencies of nonlinear Compton scattering with those of Raman scattering. In the ultra-relativistic regime the growth rate of the Raman scattering is rather low despite the enhancement caused by the radiation reaction force~\cite{Kumar_etal_2013}. However, growth of the Raman scattering in the relativistic case can be comparable with the Compton signal. Like Compton scattering, Raman scattering of linearly polarized light can exhibit harmonics of the anti-Stokes quanta of Raman scattering~\cite{Sakharov_Kirsanov_1997}. Fig.~\ref{fig4} depicts the difference between the Compton peaks from Fig.~\ref{fig3} and the anti-Stokes mode $\omega_{+}^n = n \omega_L + \omega_p, \textrm{with}\, n=1,2,3$. For the fundamental quanta, i.e., $\omega_{+}^1 =  \omega_L + \omega_p$, one can see for all plasma densities there is significant interference between these two terms. This strongly suggests that one should also expect possible quantum interference effects on the Raman spectra.  These interferences can cause broadening of the Raman signal, which is sometimes also observed in experiments due to other origins. Moreover, at high plasma densities and for higher Compton peaks, the difference gets larger. Hence in this parameter regime it is possible to differentiate between the Compton and the Raman signals. These predictions can be readily verified for a linearly polarized laser pulse of few femtosecond duration with intensity $I_L\sim 10^{22}$W/cm$^2$ and plasma densities $n_e\sim 10^{16-19}$cm$^{-3}$, which are already available. According to Fig.~\ref{fig1}, not only the total photon yields are increased but also the maximum photon energy at higher plasma densities. Also the broadening of the SRS signal with respect to plasma density should be easily detected with the current state-of-art interferometric analysis such as, e.g., SPIDER and FROG techniques~\cite{Iaconis_Walmsley_1998,OShea_etal_2001}.

To summarize, we have studied nonlinear Compton scattering of an electron in the presence of a strong few-cycle laser field modified in a background plasma based on modified Volkov states. \textcolor{black}{The found modifications of the emission spectra arising due to the laser dispersion in a plasma} suggest an impact of the plasma on conventional quantum interference effects. We discussed the implications of our results in the context of intense short-pulse laser-plasma interaction experiments and identified the quantum interference effects as an additional mechanism for the broadening of the Raman signals.


%

\end{document}